\documentclass[twoside,twocolumn]{article}
\usepackage[super,sort&compress,comma]{natbib} 
\usepackage{times,mathptmx}
\usepackage{graphicx} 
\usepackage{lastpage}
\usepackage{float}
\usepackage[english]{babel}
\addto{\captionsenglish}{%
  
}
\usepackage[T1]{fontenc}
\usepackage{mathtools} 
\usepackage[utf8]{inputenc}	
\usepackage{textcomp}	
\usepackage{hyperref} 
\usepackage{gensymb}	
\newcommand{\uvec}[1]{\boldsymbol{\hat{\textbf{#1}}}}		

\begin{document}
\twocolumn[
  \begin{@twocolumnfalse}
\noindent\LARGE{\textbf{Bridging the Gap: 3D Real-Space Characterization of Colloidal Assemblies via FIB-SEM Tomography$^\dag$}} \\ 

\noindent\large{Jessi E.S. van der Hoeven,\textit{$^{a,b,\ddag}$}$^{\ast}$ Ernest B. van der Wee,\textit{$^{a,\ddag}$}$^{\ast}$ D.A. Matthijs de Winter,\textit{$^{a}$} Michiel Hermes,\textit{$^{a}$} Yang Liu,\textit{$^{a,c}$} Jantina Fokkema,\textit{$^{a}$} Maarten Bransen,\textit{$^{a}$} Marijn A. van Huis,\textit{$^{a}$} Hans C. Gerritsen,\textit{$^{a}$} Petra E. de Jongh,\textit{$^{b}$} and Alfons van Blaaderen\textit{$^{a}$}}$^{\ast}$ \\

\vspace{0.25cm}

\small{
\textit{$^{a}$~Soft Condensed Matter and Biophysics, Debye Institute for Nanomaterials Science, Utrecht University, Princetonplein 1, 3584 CC Utrecht, The Netherlands\\
$^{b}$~Inorganic Chemistry and Catalysis, Debye Institute for Nanomaterials Science, Utrecht University, Universiteitsweg 99, 3584 CG Utrecht, The Netherlands \\
$^{c}$~Department of Earth Sciences, Utrecht University, Budapestlaan 4, 3584 CD Utrecht, The Netherlands \\
\dag~Electronic Supplementary Information (ESI) available: Movies S1-12 show the transmission electron tomography, FIB-SEM tomography and confocal microscopy data stacks and 3D reconstructions. Figures S1-4 give additional information on our image processing methods and show extra electron tomography data. See \href{https://pubs.rsc.org/en/Content/ArticleLanding/2019/NR/c8nr09753d}{DOI: 10.1039/C8NR09753D} \\
\ddag~These authors contributed equally to this work \\
$^{\ast}$E-mail: j.e.s.vanderhoeven@uu.nl, e.b.vanderwee@uu.nl, a.vanblaaderen@uu.nl }
}
\vspace{0.5cm}

 \noindent\normalsize{Insight in the structure of nanoparticle assemblies up to a single particle level is key to understand the collective properties of these assemblies, which critically depend on the individual particle positions and orientations.  
However, the characterization of large, micron sized assemblies containing small, 10-500 nanometer, sized colloids is highly challenging and cannot easily be done with the conventional light, electron or X-ray microscopy techniques.
Here, we demonstrate that focused ion beam-scanning electron microscopy (FIB-SEM) tomography in combination with image processing enables quantitative real-space studies of ordered and disordered particle assemblies too large for conventional transmission electron tomography, containing particles too small for confocal microscopy.
First, we demonstrate the high resolution structural analysis of spherical nanoparticle assemblies, containing small anisotropic gold nanoparticles. Herein, FIB-SEM tomography allows the characterization of assembly dimensions which are inaccessible to conventional transmission electron microscopy.
Next, we show that FIB-SEM tomography is capable of characterizing much larger ordered and disordered assemblies containing silica colloids with a diameter close to the resolution limit of confocal microscopes.
We determined both the position and the orientation of each individual (nano)particle in the assemblies by using recently developed particle tracking routines.
Such high precision structural information is essential in the understanding and design of the collective properties of new nanoparticle based materials and processes.}
\vspace{0.5cm}
 \end{@twocolumnfalse} \vspace{0.3cm}
]

\section*{Introduction}
The collective properties of particle ensembles are highly structure sensitive and can deviate significantly from the properties of the single nanoparticles~\cite{Nie2009,Grzelczak2010,Boles2016}. Depending on the interparticle spacing, and local and global symmetry, the plasmonic, magnetic or electronic coupling between the particles can be tuned, giving rise to altered optical, catalytic and magnetic behavior~\cite{Boles2016,Vogel2015,Montanarella2017,Hamon2016,Kang2013}.
The final 3D structures of colloidal assemblies also provide insight in the assembly process and the interactions between the colloidal particles. For example, assembled structures formed in or out of equilibrium contain information on the phase behaviour or on the glass transition or aggregation, respectively, of the colloidal particles during the assembly~\cite{Anderson2002,Gasser2001,Dinsmore2002,Montanarella2018,Kuijk2012}.

Various scattering- and microscopy techniques have been used to access the structural properties of these particle assemblies. While scattering techniques can directly probe long-range periodic order averaged over macroscopic volumes~\cite{Petukhov2015}, microscopy techniques can reveal local structures at a single particle level in real-space~\cite{Harke2008,Blaaderen1995,Friedrich2009}. Microscopy studies therefore provide insight in the presence of defects~\cite{Tilley2008,Alsayed2005}, which strongly influence the material properties and which are generally very hard to determine by scattering techniques, as these usually average over large numbers of particles and have a strong bias in detecting order over local disorder. 

Depending on the applied radiation source - X-rays, electrons or visible light - particle assemblies can be studied at different length scales, ranging from \aa ngstr\"{o}ms to micrometers. X-ray microscopy techniques enable real-space imaging of the material's local structure~\cite{Waigh2012,Byelov2013}, where the large penetration length of X-rays makes it possible to study thick and opaque colloidal assemblies in 3D~\cite{Hilhorst2012}. Nowadays, the spatial resolution of X-ray microscopy can be as precise as 10-30 nm with a sample thickness of 0.05-20 $\mu$m depending on the X-ray energy and the material properties of the sample~\cite{Groot2010}. However, the image acquisition can only be carried out at synchrotron facilities and irradiation damage can occur, especially in soft polymer based systems~\cite{Wang2009}. 
 
For a significantly higher resolution (0.1-0.5 nm) electron microscopy can be used to obtain real-space structural information. Scanning electron microscopy (SEM) allows imaging of the assembly's exterior and provides information on the surface structure, whereas transmission electron microscopy (TEM) and, in particular, transmission electron tomography in combination with particle fitting algorithms can reveal the positions and orientations of the particles in the interior of colloidal assemblies~\cite{Friedrich2009,DeNijs2015,Zanaga2016,Altantzis2017,Wang2018}. Most materials science systems analysed by transmission electron tomography are investigated in STEM-HAADF imaging mode (scanning transmission electron microscopy - high angle annular dark field), where the so-called Z-contrast stems from the difference in (high-angle) scattering power of the elements constituting the sample. When there is a sufficient difference in Z-contrast between two types of colloidal particles, 3D characterization of binary systems becomes feasible as well~\cite{Friedrich2009}. 
An important limitation in the quantitative interpretation of tomography data is the fact that it is not possible to image the object of interest over the full 180$\degree$ range. This so-called missing wedge problem causes artefacts in the reconstruction. In addition, the limited penetration depth of the electron beam in larger assemblies and high Z-contrast materials limits the maximum assembly size that can be quantitatively characterized to about 500 nm~\cite{Altantzis2017,Zanaga2016}.

Light microscopy techniques, on the other hand, can have larger penetration depths~\cite{Elliot2001}. When the sample is refractive index matched and a dye is incorporated in the particles, confocal microscopy is capable of resolving large assemblies of >500 nm colloids in 3D~\cite{Blaaderen1995,Prasad2007,Leahy2018}. The sample thickness can be up to 300 $\mu$m for high numerical aperture (NA) objectives~\cite{Martini2002}. 
The particle positions of both spherical and anisotropic particles can be extracted using multiple particle fitting and tracking algorithms~\cite{Crocker1996,Besseling2015,Dassanayake2000}. 
In order to improve the resolvability of the particles, image restoration techniques using the point spread function (PSF) of the microscope can be used~\cite{Monvel2001}.
The advent of super-resolution techniques, such as stimulated emission depletion (STED), have made it possible to image colloidal assemblies at even higher resolutions. 
The axial (\textit{Z}-direction) resolution is still limiting but has been brought down recently below 100 nm, allowing particles sizes of 200 nm to be resolved in 3D~\cite{Hell2007,Harke2008,Vicidomini2018}. 
However, STED microscopy requires better dyes and is sensitive to refractive index mismatch. 
In practice, large confocal-like volumes are not easily imaged with STED either.
This means that neither X-ray nor conventional electron nor light microscopy are able to image large sample volumes of (non-index matched) materials at a nanometer resolution. 

Focused ion beam - scanning electron microscopy (FIB-SEM) tomography does offer the unique opportunity for high resolution 3D real-space imaging of hundreds to thousands of cubic microns with a resolution down to a few nanometers~\cite{Cantoni2014}. 
FIB-SEM relies on a dual beam approach, using both a focused ion and electron beam. 
Herein, both beams usually have their own column and lens system, allowing them to operate independently. 
The FIB scans a focused beam of gallium ions onto the sample surface. 
The momentum transfer of the gallium ions results in a sputtering process called milling. 
Precision milling results in trenches at predetermined locations, allowing the SEM to record high-resolution images of sections of the material of interest. 
Consecutive slices as thin as 3 nm can be milled away by the FIB, while the SEM records high resolution images in between the milling. 
This process is called FIB-SEM tomography. 
Successful examples of FIB-SEM tomography are found in many disciplines and it has been applied to \itshape e.g. \upshape inorganic nanomaterials~\cite{Perrey2004,Hamon2016,Winter2016}, photonic crystals~\cite{Galusha2009}, biological tissue~\cite{Narayan2015,Winter2009} and porous geological materials~\cite{Yang2016,Winter2009}. 

In this work, we demonstrate the use of FIB-SEM tomography in the 3D characterization of colloidal assemblies with nano- to micrometer sized colloidal particles. 
We show that for assemblies of gold nanorods, TEM tomography is limited to assemblies composed of less than 100 particles, whereas FIB-SEM tomography can be used to characterize assemblies of more than 1000 particles. 
In addition, we show the use of FIB-SEM tomography in the structural analysis of disordered and ordered assemblies composed of single and binary species of $\sim$0.5 $\mu$m sized silica particles. We compare this to confocal microscopy in combination with image restoration and discuss the advantages of FIB-SEM tomography.

\section*{Results}
\subsection*{FIB-SEM tomography for particle assemblies}
\begin{figure*}[h]
\centering
  \includegraphics[width=0.8\textwidth]{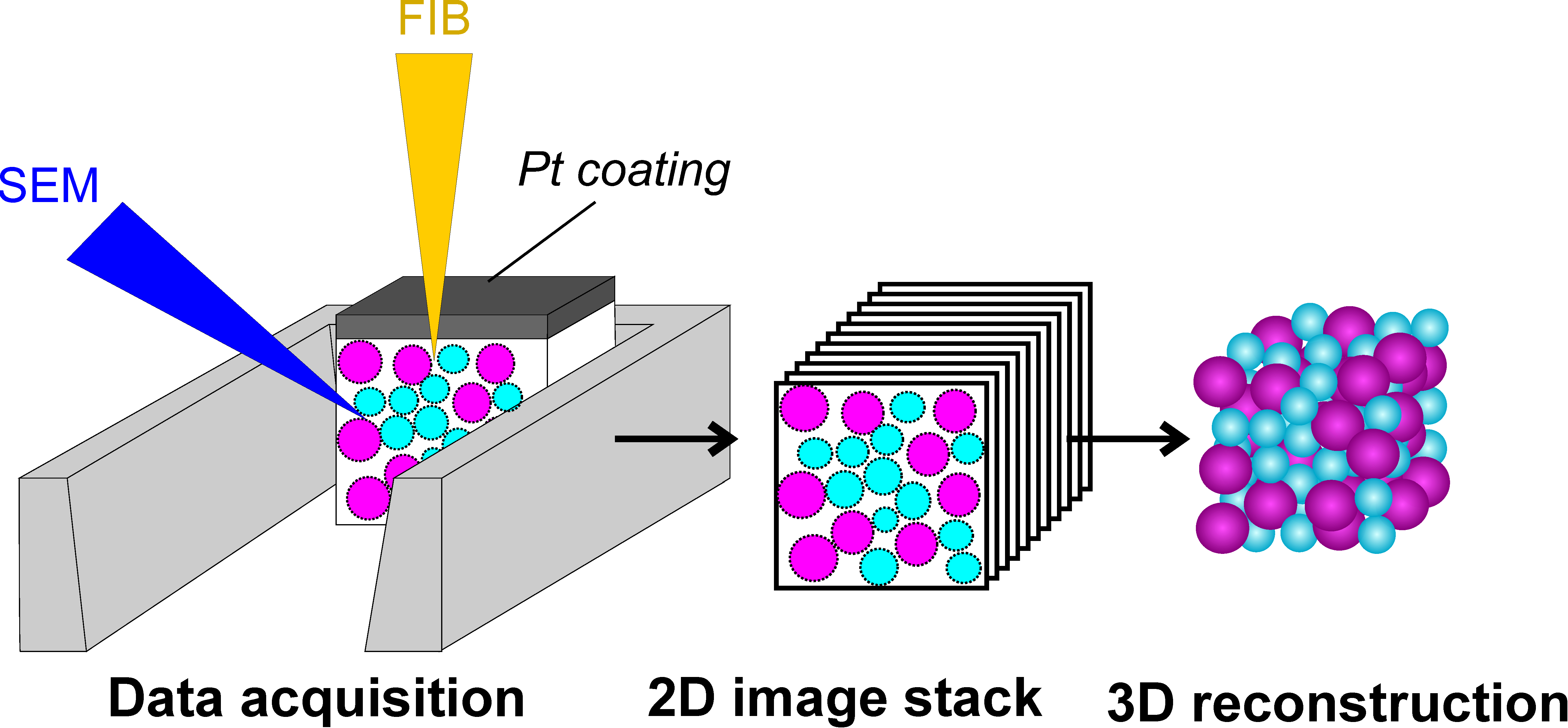}
  \caption{\textbf{3D characterization of colloidal assemblies with FIB-SEM tomography.} Left: the tomography data acquisition, obtained by iteratively removing a slice of the assembly with the FIB beam (yellow) and imaging of the assembly with the electron beam (dark blue). Middle: the obtained stack of 2D images acquired at different $Z$-depths. Right: 3D reconstruction of the particle coordinates from the 2D image stack.}
  \label{fig:Figure1_FIB-SEM}
\end{figure*}

We applied FIB-SEM tomography to three 3D assemblies: a $<$1~$\mu$m$^3$ sized nanoparticle (NP) assembly, consisting of silica coated gold nanorods (AuNRs, $l_{Au}$ = 119 nm (11\% PDI), $d_{Au}$ = 16 nm (13\% PDI)), a much larger $\sim$1,000~$\mu$m$^3$ sized assembly composed of large spherical silica colloids ($d$ = 531 nm, <2\% PDI) and a similarly sized assembly composed of a binary glass of the same spheres mixed with smaller silica spheres ($d$ = 396 nm, 1\% PDI).
In Figure~\ref{fig:Figure1_FIB-SEM} we depict the general approach in which FIB-SEM tomography is used in the 3D characterization of particle assemblies.
The characterization can be divided in three stages: (1) acquisition of the tomography series, (2) alignment of the 2D image stack and (3) fitting of the positions and orientations of the particles in 3D. 

We used two different ways of sample preparation depending on the type and size of the particle assembly. For the large colloidal assemblies consisting of the 531 nm silica spheres, the assembly was embedded in a resin, to preserve the assembly structure during the milling process by the FIB. This is essential to correctly determine the initial particle coordinates and orientations. Thereafter, a conductive platinum layer was sputter coated on top of the ensemble at the region of interest to prevent charging during FIB-milling and/or SEM-imaging. The small spherical AuNR nanoparticle assemblies, called supraparticles, were not embedded in a resin, but the selected supraparticle was only covered with a Pt-coating, which prevented both charging and deformation of the spherical assembly shape during the milling process. In the tomography data acquisition the slice thickness was varied for the different colloidal particle sizes and was chosen such that at least 6 slices through each individual particle were obtained. Thereafter, the SEM images were aligned and the coordinates and orientation of the individual particles determined. For particle identification of the nanorod assemblies we used the rod-tracking code developed by Besseling $et$ $al.$ ~\cite{Besseling2015}. For the micron sized colloidal assemblies we used a more recent analysis method in which the particles are identified with gradient tracking. 
The gradient based tracking approach is a more general method in comparison to the rod-tracking code which can only be applied to rod-like particles~\cite{Besseling2015}. Figure S1 in the supporting information outlines the main principles of gradient based tracking. 
 
\subsection*{High resolution 3D imaging of gold nanorod assemblies}
\begin{figure}[h!]
	\centering
	\includegraphics[width=0.45\textwidth]{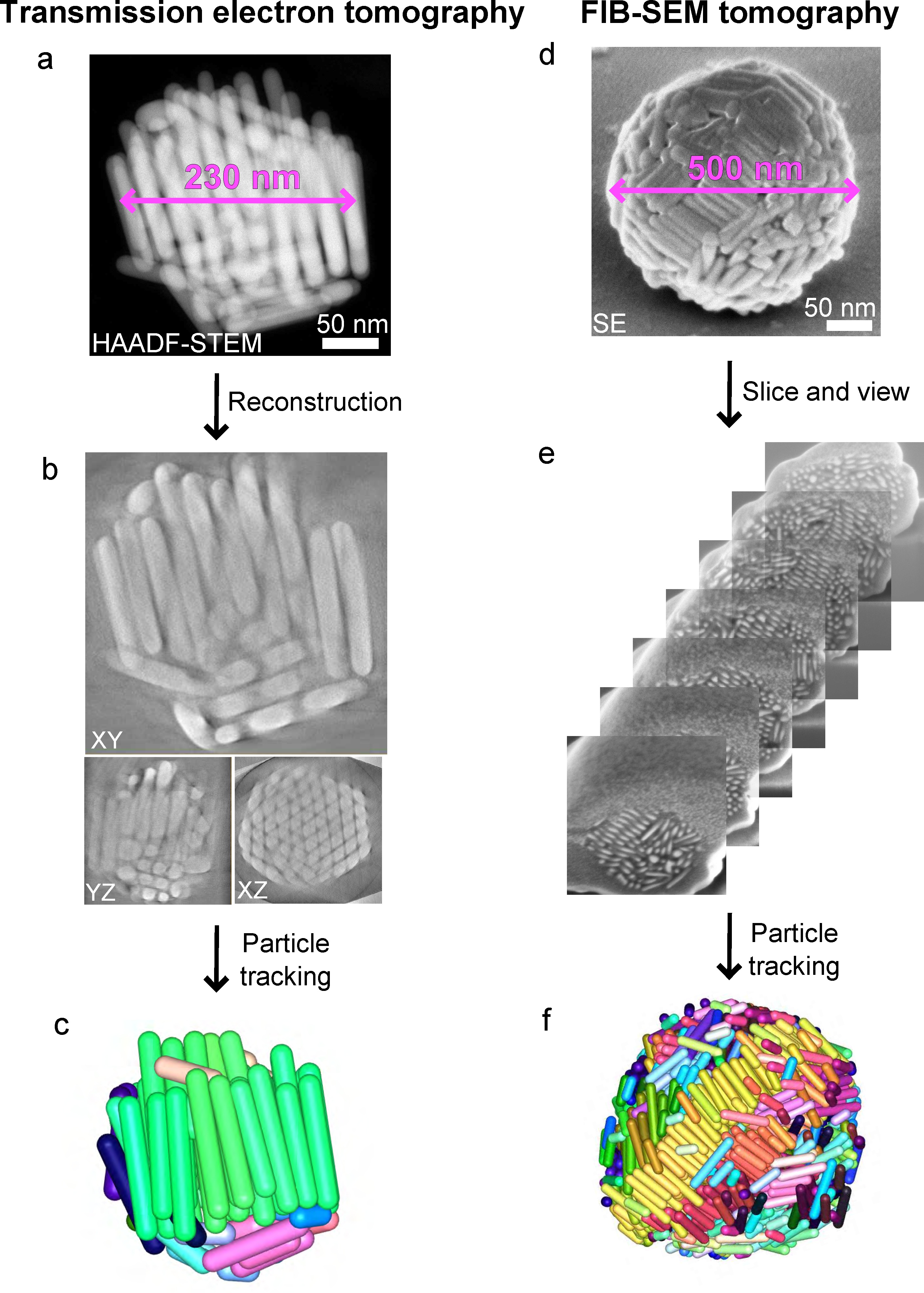}
	\caption{\textbf{3D characterization of differently sized silica coated gold nanorod assemblies with transmission electron and FIB-SEM tomography.} Left: transmission electron tomography of a small AuNRs@SiO$_2$ ($l_{Au}$ = 119~nm (11\% PDI), $d_{Au}$ = 16~nm (13\% PDI)) assembly with $d = $~230~nm, consisting of 96 nanorods: a) Single HAADF-STEM image, acquired at 0$\degree$ tilt, b) $XY$, $YZ$ and $XZ$ orthoslices of the assembly's interior, after reconstruction of the tilt series, c) tracking of the position and orientation of the nanorods in 3D, where the rods are colored according to their orientation. Right: FIB-SEM tomography of a larger AuNRs@SiO$_2$ assembly with $d = 500$~nm, consisting of 1,279 nanorods: d) SE-image of the exterior of the AuNRs@SiO$_2$ assembly,  e) SE images acquired while milling into the interior of the assembly with the FIB, f) 3D representation of the tracked AuNRs in the assembly.}  \label{fig:Figure2_AuNRs}
\end{figure}

For the FIB-SEM tomography on nanoparticle based assemblies, we prepared $\sim$200~nm to 2~$\mu$m large spherical supraparticles of silica coated gold nanorods ($l_{Au}$~=~119~nm (11\%~PDI), $d_{Au}$~=~16~nm (13\%~PDI)). This type of nanoparticle system is particularly interesting for Raman spectroscopy, where the Raman enhancement depends on the overlap between the surface plasmons of the individual gold particles and thus on the precise position and orientation of the nanorods~\cite{Hamon2016}. 
To obtain the AuNR assemblies, we first synthesized colloidal gold nanorods~\cite{Ye2013} coated with a 3 nm thin silica shell, functionalized with a hydrophobic coating~\cite{Liu2014}. 
Subsequently, the rods were assembled in spherical clusters by using a solvent evaporation method~\cite{DeNijs2015} that we recently also applied to rod-like particles~\cite{Besseling2015} (see Experimental section for more synthesis details). 

We applied both transmission electron tomography and FIB-SEM tomography to obtain the 3D structure of the AuNR assemblies. In Figure~\ref{fig:Figure2_AuNRs} we show the transmission electron and FIB-SEM tomography results for the characterization of a small and a larger AuNR supraparticle, of which the spherical shape is usually well suited for transmission electron tomography~\cite{DeNijs2015,Altantzis2017,Wang2018}. Figure \ref{fig:Figure2_AuNRs}a-c shows the tilt series, reconstruction and 3D model of a 230~nm assembly obtained \itshape via \upshape transmission electron tomography. In the 3D model the rods are color-coded based on their orientation, showing that the rods are preferentially ordered in the same direction. For this relatively small assembly the positions and orientations of all 96 rods could successfully be obtained from the 3D reconstruction. The transmission electron tomography tilt series, reconstruction and 3D model of the tracked AuNR assembly can be viewed in Movie S1-S3, respectively.

Due to the limited penetration depth of the electron beam caused by the high Z-contrast of the Au atoms, transmission electron tomography can only be applied to small particle assemblies for this type of systems. To illustrate this we performed transmission electron tomography on a larger, 340~nm ensemble composed of the same AuNRs as the assembly shown in Figure \ref{fig:Figure2_AuNRs}a-c. In Figure S2 we show that the 340~nm assembly was too large to obtain a high quality reconstruction. To access the full structural properties of larger and/or denser assemblies, we applied FIB-SEM tomography. In Figure~\ref{fig:Figure2_AuNRs}d-f we show the secondary electron (SE) image of the exterior, part of the FIB-SEM tomography series of the interior and the 3D reconstruction of a 500~nm AuNR supraparticle, consisting of the same AuNRs as the assembly in Figure \ref{fig:Figure2_AuNRs}a-c. In order to reliably distinguish the individual NRs, the lowest possible $Z$ step size of 3 nm had to be used, such that at least 6 slices per rod were obtained. The tomography series consisted of 160 $XY$-slices (2304 $\times$ 2048 pixels), spaced 3~nm apart resulting in a voxel size $X\times Y\times Z$ of 0.3244 $\times$ 0.411 $\times$ 3 nm$^3$. The total imaged volume was 0.300 $\mu$m$^3$. The particles coordinates and orientations were determined by making use of a rod fitting algorithm~\cite{Besseling2015}.
From the FIB-SEM tomography data set in Figure \ref{fig:Figure2_AuNRs}e we obtained the positions and orientations of 1,279 rods. The complete FIB-SEM tomography series and 3D model of the tracked AuNR assembly can be found in Movie S4 and S5, respectively.

\subsection*{FIB-SEM tomography of a colloidal crystal}
To demonstrate the feasibility of FIB-SEM tomography to also analyze much larger colloidal assemblies containing micron sized particles, we prepared a colloidal crystal of monodisperse silica spheres ($d = 531$ nm, <2\% PDI). 
About one percent of the particles had a 30 nm gold core, whereas the other 99 percent had a 45 nm fluorescently (rhodamine B isothiocyanate, RITC) labeled core to also enable characterization with confocal microscopy. 
The crystal was grown by controlled vertical deposition at elevated temperature onto a glass slide~\cite{Jiang1999}, resulting in a thickness of $\sim$11 $\mu$m (which corresponds to $\sim$25 layers).

 \begin{figure*} [!h]
	\centering
	\includegraphics[width=0.7\textwidth]{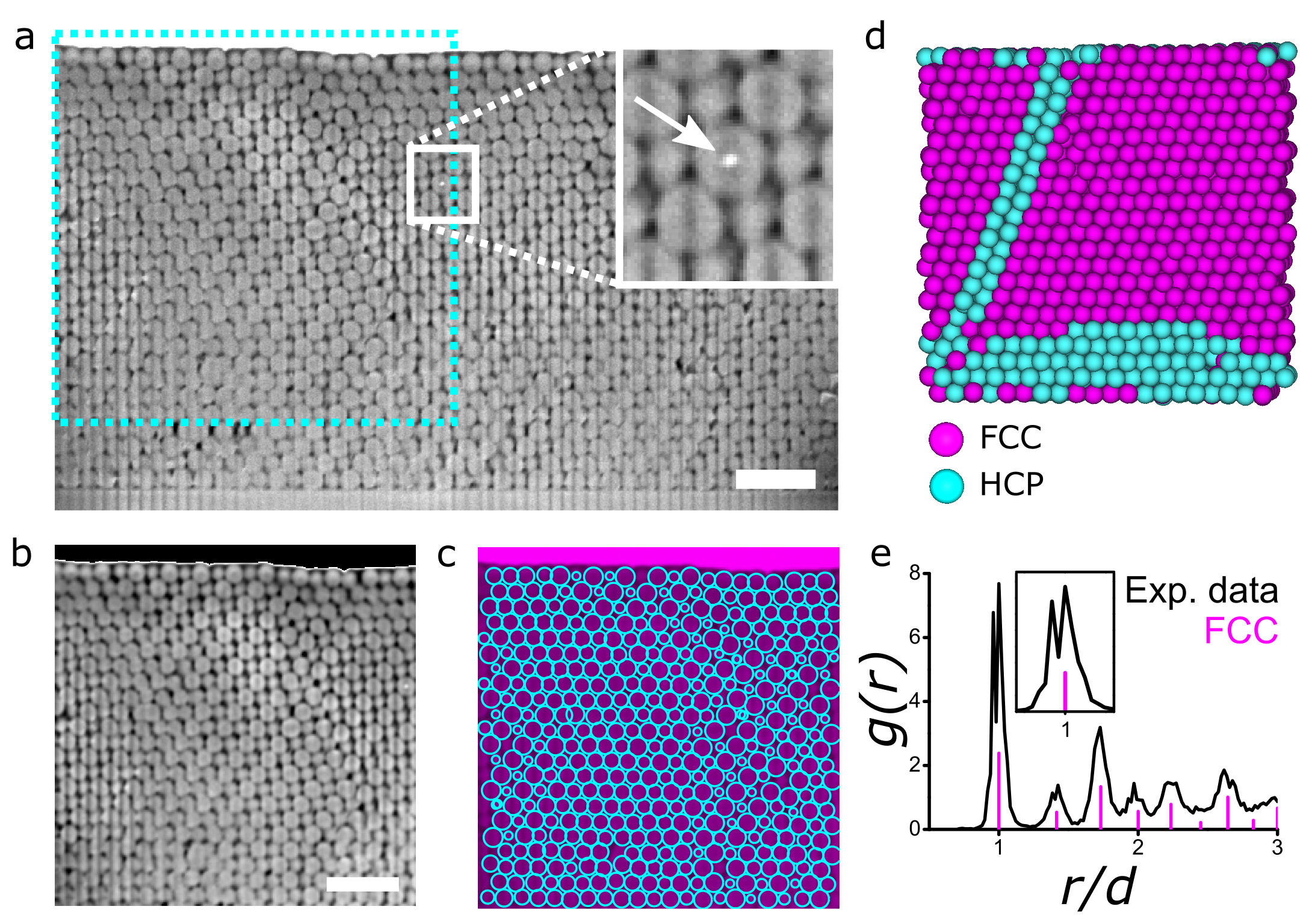}
	\caption{\textbf{FIB-SEM tomography on a crystal of silica colloids ($d$~= 531 nm, <2\% PDI).}
a) Slice from FIB-SEM tomogram with a total volume of 2,610 $\mu$m$^3$. Arrow in inset points at the gold core of a particle.
b) Zoom-in of the dashed cyan rectangle in the (a).
c) Overlay of (b) with cyan circles indicating identified particles.
d) Cut-through of computer rendering of coordinates from the reconstruction in (c) with colors of particles assigned to local symmetry of particles as calculated with bond orientational order parameters showing that the crystal structure is majorly FCC (magenta) with a horizontal stacking fault at the bottom and a slanted stacking fault running through the structure, both with HCP symmetry (cyan). The reconstructed volume is 1,000 $\mu$m$^3$, with 8,891 particles.
e) Radial distribution function $g(r)$ calculated from coordinates of the rendering partly shown in (d) (black), compared to the peaks of an ideal FCC crystal (magenta). The inset shows the double peak in the $g(r)$ at $r / d \approx 1$, due to the shrinkage in the growth direction of the crystal.
The scale bars are 2 $\mu$m.
}  \label{fig:figure_3_xtal}
\end{figure*}

In Figure~\ref{fig:figure_3_xtal}a we show a slice from the FIB-SEM tomogram with a pixel size in \textit{X} and \textit{Y} of 10.5 nm, recorded with a milling step size in \textit{Z} of 50.0 nm. The total sampled volume was 2,610 $\mu$m$^3$.  
The inset in Figure~\ref{fig:figure_3_xtal}a shows a gold core in one of the silica particles, demonstrating the possibility of investigating multiple length scales in hierarchical assemblies using FIB-SEM tomography. From the full data stack we cropped a volume of 1,000 $\mu$m$^3$ (dashed cyan rectangle in Figure~\ref{fig:figure_3_xtal}a) for reconstruction, see Figure~\ref{fig:figure_3_xtal}b. Using a gradient tracking algorithm 
the particle coordinates were obtained, as we show in Figure~\ref{fig:figure_3_xtal}c where the particles positions are depicted by the cyan circles. In Movie S6 and S7 the full FIB-SEM tomography series and corresponding 3D model, respectively, are shown.

To obtain insight into the structure of the crystal, we calculated the local bond orientational order of every particle in the assembly~\cite{Lechner2008}. In Figure \ref{fig:figure_3_xtal}d we show a computer rendering of the particle assembly, where the particles are colored according to their local symmetry (see Experimental Section for details). Although the majority of the particles have local face-centered cubic (FCC) symmetry, the particles at the bottom of the reconstructed volume are packed locally with hexagonal close-packed (HCP) symmetry. Moreover, a slanted stacking fault runs through the crystal, also with local HCP symmetry.
When the radial distribution function $g(r)$ is calculated from the reconstructed coordinates (8912 particles), a good agreement with the FCC structure is found (see Figure~\ref{fig:figure_3_xtal}e). There is however a double peak at $r / d \approx 1$, which is absent in close packed crystals grown in bulk or by gravity~\cite{Hoogenboom2003}. From the ratio of the $r / d $ values of the two peaks in Figure~\ref{fig:figure_3_xtal}e it follows that the difference is close to 4\%.  This is in good agreement with previous work on colloidal crystals grown using the vertical deposition method, where the same $\sim$4\% of shrinkage in the growth direction in the hexagonal (\textit{111}) planes has been measured with X-ray diffraction and confocal microscopy~\cite{Thijssen2007}. 

\subsection*{Characterization of a binary colloidal glass}
FIB-SEM tomography can also be used to obtain real-space information of binary particle systems. Here we intentionally made a binary glassy sample as it is more difficult to retrieve the particle coordinates from the microscopy data in comparison to a crystalline structure. To demonstrate this, we mixed the previously used 531 nm RITC labeled silica colloids with smaller 396 nm (1\% PDI) silica particles, which had a fluorescently (fluorescein isothiocyanate, FITC) labeled core of $\sim$200 nm. For comparison, the particles were imaged with both confocal laser scanning microscopy and FIB-SEM tomography. 

\begin{figure*} [!h]
	\centering
	\includegraphics[width=0.7\textwidth]{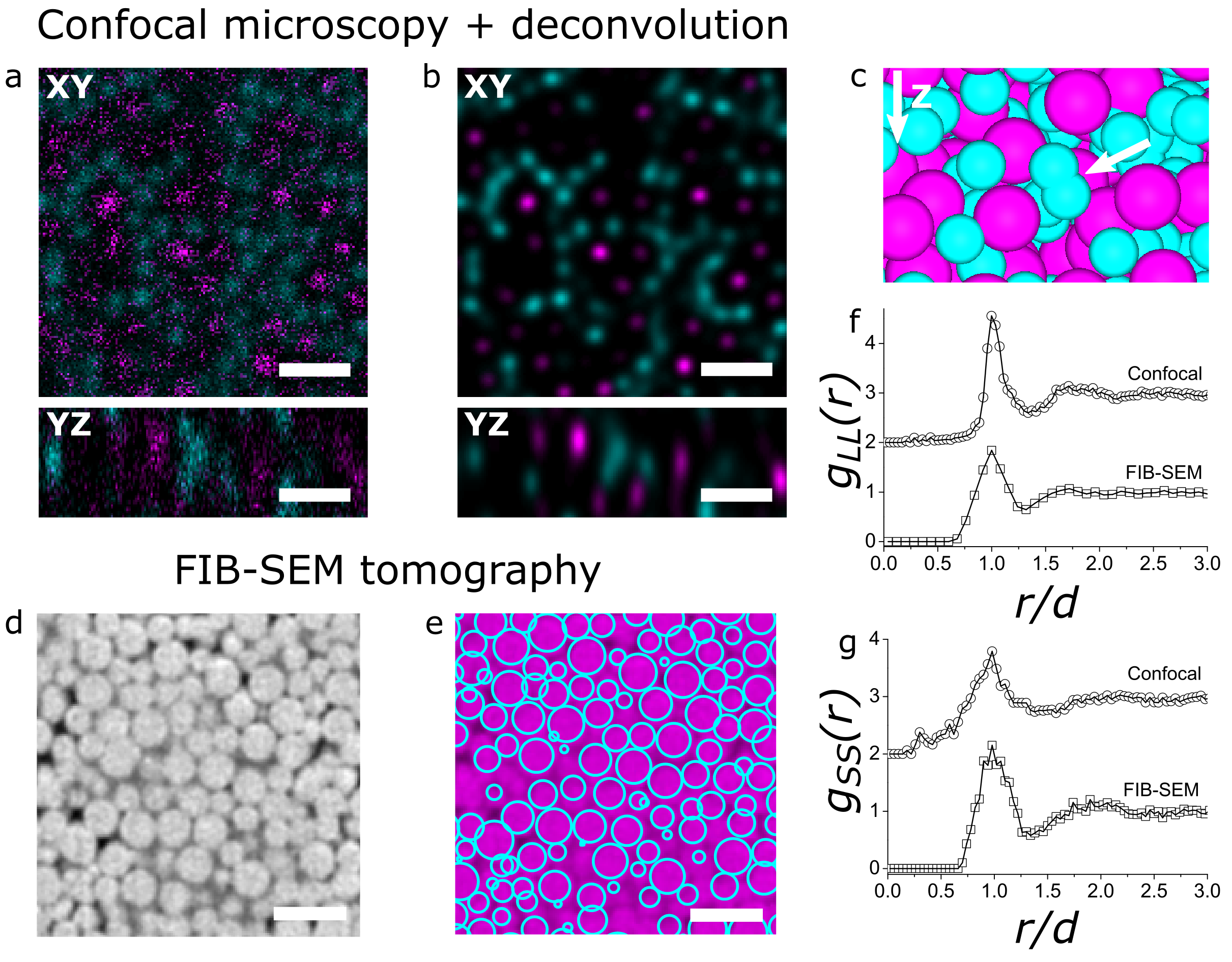}
	\caption{
\textbf{Binary glass characterized by confocal microscopy, in combination with image restoration, and FIB-SEM tomography.}
a) \textit{XY} and \textit{YZ} slices from a two channel confocal microscopy image stack of a binary glass of 396 nm (1\% PDI) fluorescein (cyan, S) and 531 nm (<2\% PDI) rhodamine (magenta, L) labeled core-shell silica colloids, with a total volume of $\sim$1,200 $\mu$m$^3$.
b) Same slices after deconvolution of the image stack.
c) Fragment of computer rendering of coordinates identified from the image stack in b). 
The arrow points at two overlapping particles, where the particle tracking algorithm misidentified two particles with a small separation in the axial direction.
d) Fragment of FIB-SEM tomogram of the same binary glass with a total volume of $\sim$500 $\mu$m$^3$.
e) Overlay of d) with cyan circles indicating identified particles.
Partial radial distribution functions $g_{LL}(r)$ (f) and $g_{SS}(r)$ (g) from the coordinates obtained through confocal microscopy and image restoration, and FIB-SEM tomography.
The scale bars are 1 $\mu$m.
}  \label{fig:Figure4_binaryglass}
\end{figure*}

For confocal microscopy, the particles were drop casted from an ethanol dispersion on a cover glass and refractive index matched with a mixture of glycerol and $n$-butanol ($n_D^{23}$~=~1.44). Image stacks of the two differently labelled particles were imaged sequentially, as shown in Figure~\ref{fig:Figure4_binaryglass}a, spanning a volume of $\sim$1,200 $\mu$m$^3$. Figure~\ref{fig:Figure4_binaryglass}b shows the stacks after image restoration, which involves deconvolution of the data with the microscope point spread function using the Huygens (SVI) deconvolution software. The deconvoluted confocal data stack of the binary glass can be viewed in Movie S8. Using a classical particle tracking routine~\cite{Crocker1996} extended to 3D data sets~\cite{Blaaderen1995}, we identified the coordinates of both species in the assembly. A fragment of a computer rendering of the coordinates is shown in Figure~\ref{fig:Figure4_binaryglass}c (the full set of coordinates can be viewed in Movie S9), from which the partial radial distribution functions of the large ($g_{LL}(r)$, 4192 particles) and small particles ($g_{SS}(r)$, 6544 particles) were calculated (Figure~\ref{fig:Figure4_binaryglass}f,g).  

For FIB-SEM tomography, the particles were embedded in a resin after dropcasting. A stack with a total volume of $\sim$1,000 $\mu$m$^3$ was recorded with a FIB milling step size of 50 nm. From this stack, a volume of $\sim$500 $\mu$m$^3$ was cropped for particle identification (Figure~\ref{fig:Figure4_binaryglass}d). The coordinates of the particles were obtained using a gradient tracking algorithm,
 where the particle sizes were fitted for every particle (Figure~\ref{fig:Figure4_binaryglass}e). This resulted in a distribution of sizes with two peaks where the population was divided into small and large species using a threshold diameter of 475 nm. 
From the coordinates of the different particles, the partial radial distribution functions $g_{LL}(r)$ (2448 particles) and $g_{SS}(r)$ (2817 particles) were calculated, as shown in Figure~\ref{fig:Figure4_binaryglass}f and g, respectively. Movie S10 and S11 show the FIB-SEM tomography series and the corresponding 3D model of the binary glass.

When comparing the partial radial distribution functions of the large ($g_{LL}(r)$) and small spheres ($g_{SS}(r)$) acquired using the two techniques, an agreement was found for the peak positions in the $g_{LL}(r)$, although the $g_{LL}(r)$ from FIB-SEM had a broader first peak (Figure~\ref{fig:Figure4_binaryglass}f). The functions of the smaller particles $g_{SS}(r)$, however, disagreed to a higher extend (Figure~\ref{fig:Figure4_binaryglass}g). The radial distribution function calculated from the coordinates obtained by confocal microscopy had a lower first peak and was non-zero at values smaller than the smallest distance the particles can be apart ($\sim$390 nm). This points at overlapping particles due to mis-identification of the smaller particles positioned relatively close to each other in the axial direction of the confocal microscope, as reported in Ref.~\cite{Leocmach2013}.  An example of such overlapping particles in the computer rendering of the coordinates is shown in Figure~\ref{fig:Figure4_binaryglass}c. These types of errors were absent in the confocal $g_{LL}(r)$, indicating that for the small particles, the limit of the (axial) resolving power of the confocal microscope was approached. FIB-SEM tomography, on the other hand, does have sufficient resolving power to identify the positions of the smaller particles correctly.

\section*{Discussion}
\subsection*{Data acquisition}
During the FIB-SEM data acquisition it is crucial that the colloidal particles are imaged in their original positions and orientations within the ensemble. Depending on the type of  assembly various changes in the ensemble structure can occur during the tomography. 
Supraparticles, especially composed of NPs, are prone to deform to a more flat, non-spherical structure during FIB exposure and should therefore be encapsulated in a Pt coating before tomography. 
On the other hand, in the image acquisition of the assemblies composed of the micron sized colloids we noticed that particles could "fall off" during the milling process, when the particles are no longer supported by their neighbors. 
This can cause a shift in the apparent position of the particles in the 3D reconstruction. To prevent this, it is advisable embed the particle assembly in a resin prior to the image acquisition.

When imaging porous assemblies with FIB-SEM, so-called curtaining effects are likely to arise due to the different (material) densities. 
Curtaining occurs when the milling speed in the region of interest is inhomogeneous, resulting in different slice thicknesses in the milling direction. 
Such inhomogeneities in slice thickness complicate or even prohibit a quantitative reconstruction of the correct assembly structure in 3D. 
We observed these curtaining effects when milling the relatively porous and thick colloidal crystal and binary glass, but not for the densely packed and thin AuNR assemblies. 
The curtaining during the data acquisition can successfully be suppressed by embedding the colloidal assemblies in a resin beforehand. 
In this way, the pores in between the particles are filled, making the milling speeds more homogeneous. 
The remaining curtaining "stripes" can be filtered out during the data processing by using fast Fourier transform (FFT) filtering. 
Herein, one calculates the FFT of the acquired image, removes the lines in the FFT patterns caused by the curtaining and performs an inverse FFT to obtain the filtered image (Figure S3).
The curtaining effect can also be suppressed using advanced acquisition methods and image processing~\cite{Loeber2017}.

Another difficulty encountered during acquisition is the accumulation of charge in the sample due to the scanning electron beam, resulting in white areas in the SEM images. Although the samples were connected to the SEM stub with conductive carbon tape and sputter coated with Pt to prevent the build-up of charge, charging still occurred. One way to reduce this effect was to acquire the SEM images at a lower beam current, and compensate for the signal reduction by integrating multiple images. Instead of modifying the acquisition parameters, the effects of charging can also be suppressed by image processing~\cite{Loeber2017}.

\subsection*{Determining particle coordinates and orientations}
There are several advantages in using the gradient based particle tracking algorithm used in this work. 
First, it is not limited to the recognition of spherical particles only, but can also be applied to different (anisotropic) shapes,~\cite{Wang2018} and therefore to a wide variety of particle assemblies. 
Second, it enables the determination of the particle orientation for each \textit{individual} particle. The ability to exactly determine the orientation and position of each NP and all interparticle distances is crucial in, for instance, calculating the assembly's collective plasmonic properties. 
Previously, only average orientations of several particles per assembly volume
could be obtained~\cite{Hamon2016}. With our particle specific analysis method it now becomes feasible to directly compare the theoretical and experimental behavior of plasmonic particle assemblies and to predict their performance for \itshape e.g. \upshape Raman spectroscopy, which is strongly influenced by the exact particle locations and the presence of so-called hot-spots, where locally electromagnetic fields can non-linearly enhance each other. 

When the contrast between the particles and their surroundings is low, tracking is more difficult. For the spherical AuNR assembly in Figure \ref{fig:Figure2_AuNRs}, the contrast between the Au of the NRs and the Pt of the protective coating was very low. Particles in or close to the Pt coating were prone to misidentification and difficult to distinguish from real particles (Figure S4). Reliable tracking was therefore only possible for the layers below the particle layer that was closest to the Pt coating. 

\subsection*{Comparing the real-space microscopy techniques}
We studied the AuNRs assemblies with both transmission electron tomography and FIB-SEM tomography. Which method is to be preferred predominately depends on the size and Z-contrast of the individual nanoparticles, and the size of the total ensemble.
Generally, the spatial resolution of the transmission electron microscope is superior to the resolution of the electron beam used in FIB-SEM tomography. More importantly, the resolution in the Z-direction for the current generation of high-end Ga-based FIB-SEM microscopes is limited to 3 nm, which is the minimum slice thickness that can be milled with the FIB. 
Since a minimum of about 6 slices per nanoparticle is required to reliably determine its position and orientation, FIB-SEM tomography is presently only suited for assemblies consisting of $\geq$18 nm particles. Although the accuracy of the tracking is generally higher than the resolution of the FIB-SEM images~\cite{Besseling2015,Crocker1996}, for now transmission electron tomography is still the preferred analysis technique for small nanoparticle assemblies.

However, for assemblies with a thickness larger than 300 nm and/or composed of high Z-contrast materials, transmission electron tomography is no longer applicable. When imaging such assemblies with transmission electron tomography, the intensity of the particles in the interior is underestimated with respect to the particles at the exterior of the assembly. This is caused by partial absorption and scattering of the incoming electron beam before reaching the inside of the particle ensemble. Likewise, the electrons that are scattered from the inside of the assembly have to penetrate a considerable amount of material before reaching the detector. This results in thickness dependent, non-linear damping of the recorded intensities, which is called a cupping artefact~\cite{Broek2012}. In the reconstruction the cupping artifact hampers a quantitative 3D structural analysis of the particle ensembles interior. An example of the cupping artifact is for example already visible in the reconstruction of the 340 nm AuNR assembly in Figure S2. Apart from post reconstruction methods to correct for the cupping effect, an alternative method to study the interior of nanoparticle assemblies larger than 300 nm is to perform microtomy prior to the transmission electron tomography measurement. Herein, one embeds the particle assemblies in a resin and cuts the sample with a diamond knife to slices as thin as 50 nm, after which electron tomography can be performed on a single slice. However, this method does not allow the continuous spatial analysis of a full particle assembly. Thus, to characterize a complete nanoparticle ensemble larger than 300 nm, FIB-SEM tomography is indispensable. 

We also compared FIB-SEM tomography to confocal microscopy for particle ensembles consisting of particles with a size close to the resolution limit of conventional confocal microscopy. For the binary glass (Figure~\ref{fig:Figure4_binaryglass}), we observed that the large spheres could still be resolved with confocal microscopy, but the smaller ($d = 396$~nm) particles could not. The overlapping particles shown in Figure~\ref{fig:Figure4_binaryglass}c indicate that the limit of the resolving power of the confocal microscope was reached. Despite the fact that more advanced particle fitting algorithms have been developed to increase the accuracy of particles position determination, these algorithms do not significantly lower the size limit of the smallest particles that can be imaged with confocal microscopy~\cite{Leahy2018,Jenkins2008,Leocmach2013,Jensen2016}. By using STED one could improve both the axial and lateral resolutions significantly (even to below 100 nm), but this technique is complicated in large sample volumes and sensitive to refractive index mismatches. FIB-SEM tomography, however, is capable of quantitatively characterizing (binary) assemblies of particles too small for confocal microscopy, without the need of refractive index matching or the incorporation of dyes in the particles. 

\subsection*{Possible future applications of FIB-SEM tomography on colloidal systems}
In this study, the assemblies were composed of particles similar in size and composition. However, the high resolution of FIB-SEM tomography would also allow the study of mixed assemblies with particle sizes ranging from 20 to 1000 nm. Either by size or by the difference in material density, different particles types can be distinguished within a mixed assembly. For example, in the case of the micron size colloidal crystal, a fraction of the silica spheres contained a much smaller (30 nm), higher density gold core instead of a silica core. The gold core could be identified in the FIB-SEM image series due to its higher Z-contrast and smaller particle size (Figure~\ref{fig:figure_3_xtal}a (inset)). In future research, FIB-SEM could thus be applied to fully characterize heterogeneous assemblies, \itshape e.g. \upshape photonic crystals composed of particles with strongly scattering cores.

The imaging method described in this work can also be applied to study low density colloidal dispersions.  To do so, the colloidal dispersions would have to be arrested prior to the imaging process. This can be done either by cryogenic quenching~\cite{Elbers2016} or chemical arrest by the polymerization of the continuous fluid phase. The latter technique enables a controlled timing of the arrest and would therefore allow the study of the different stages in assembly processes. Structural analysis of particle dispersions is also relevant in measuring for example the interparticle interactions, through the calculation of the radial distribution function. The high resolution of the FIB-SEM microscope would make it possible to start investigating interparticle interactions between nanoparticles, too small to be imaged with confocal microscopy.

\section*{Conclusions}
We have demonstrated a general approach using FIB-SEM tomography for the 3D real-space characterization of colloidal particle assemblies. 
We showed that this technique combines high resolution imaging with large sampling volumes, allowing the precise characterization of assemblies too large for conventional electron tomography, and containing particles too small to resolve with confocal microscopy. 
To this end, we first demonstrated the use of FIB-SEM tomography for high resolution imaging of nanorod assemblies. 
In contrast to conventional electron tomography, the position and orientation of the individual nanorods in assemblies larger than 300 nm could still be obtained. 
Next, we applied FIB-SEM tomography for the imaging of a colloidal crystal and a binary glass consisting of fluorescently labeled sub-micron silica spheres for large sampling volumes ($\geq$1000 $\mu$m$^3$). 
While FIB-SEM tomography was able to identify all particles in the binary glass, conventional confocal microscopy could not resolve all particles in the axial direction. 
Additionally, FIB-SEM tomography does not require the incorporation of dye in the particles or refractive index matching. 
For the data analysis we used a recently developed gradient based tracking algorithm, which can be used for different particle shapes and materials. 
In combination with such a data analysis methodology, we have shown that FIB-SEM tomography is applicable to a broad range of materials, and particle sizes and shapes, bridging and extending several other quantitative imaging techniques. 

\section*{Methods}
\subsection*{Chemicals}
All chemicals were used as received without further purification. 
Hexadecyltrimethylammonium bromide (CTAB, >98.0\%) and sodium oleate (NaOL, >97.0\%) were purchased from TCI America. 
Hydrogen tetrachloroaurate trihydrate (HAuCl$_4 \cdot$3H$_2$O) and sodium hydroxide (98\%) were purchased from Acros Organics. 
Butylamine (99.5\%), L-Ascorbic Acid (BioXtra, $\geq$99\%), cyclohexane ($\geq$99.8\%), dextran
(average molecular weight 1,500,000–2,800,000), hydrochloric acid (HCl, 37 wt\% in water),  octadecyltrimethoxysilane (OTMS, 90\%), silver nitrate (AgNO$_3$, $\geq$99\%), sodium borohydride (NaBH$_4$, 99\%), sodium silicate solution
($\geq$27\% SiO$_2$ basis, Purum $\geq$10\% NaOH), tetraethyl orthosilicate (TEOS, 98\%), sodium dodecyl sulfate (SDS $\geq$99\%), sodium citrate tribasic dehydrate, polyvinylpyrrolidone (PVP, M$_w$=10,000 g mol$^{-1}$), rhodamine B isothiocyanate (RITC, mixed isomers), (3-aminopropyl)triethoxysilane (APTES, 99\%), Igepal CO-520, ammonium hydroxide solution (ACS reagent, 28.0-30.0\% NH$_3$ basis) and $N$,$N$-dimethylformamide (DMF) were purchased from Sigma-Aldrich. 
Absolute ethanol was purchased from Merck.
Ultrapure water (Millipore Milli-Q grade) with a resistivity of 18.2 M$\Omega$ was used in all of the experiments. 
All glassware for the AuNR and gold core synthesis was cleaned with fresh aqua regia (HCl/HNO$_3$ in a 3:1 volume ratio), rinsed with large amounts of water and dried at 100 \textdegree C before use.

\subsection*{Synthesis of silica coated gold nanorod assemblies}

The preparation of the gold nanorod based assemblies consisted of four steps: colloidal synthesis of high aspect ratio AuNRs (I), silica coating (II), OTMS coating (III) and self-assembly into spherical ensembles (IV). 

The synthesis of high aspect ratio AuNRs was done according to the procedure by Ye $et$ $al.$~\cite{Ye2013}. The growth mixture consisted of CTAB (7.0 g), sodium oleate (1.23 g), Milli-Q (MQ) H$_2$O (250 mL), AgNO$_3$ (9.6 mL, 10 m\textsc{m}), HAuCl$_4$ (250 mL, 1.0 m\textsc{M}), HCl (37\%, 4.8 mL), ascorbic acid (1.25 mL, 0.064 \textsc{m}) and gold seeds (0.40 mL). The seed solution was prepared by adding an icecold NaBH$_4$ in H$_2$O solution (1.0 mL, 0.0060 \textsc{m}) to a mix of CTAB (10 mL, 0.10 \textsc{m}) and HAuCl$_4$ aqueous solution (51 $\mu$L, 50 m\textsc{m}). The resulting rods were centrifuged for 25 min at 8,000 g, washed with water and re-dispersed in 30 mL 5.0 m\textsc{m} CTAB water ($\lambda_{LSPR}$~=~1,250 nm, Ext~=~4.8, $\sim$40 mg L$^{-1}$). The resulting AuNRs had a length of 119~nm~(11\%~PDI, TEM) and diameter of 16~nm~(13\%~PDI, TEM).

The thin silica coating was carried out as follows: to the AuNRs (1.0 mL, $\lambda_{LSPR}$~=~1,250 nm, Ext~=~4.8) sodium silicate (0.15 mL, 0.54 wt\% SiO$_2$) was added while stirring vigorously. The mixture was stirred for 45 minutes at room temperature after which the rods were washed with water and ethanol, and re-dispersed in ethanol (200 $\mu$L, [Au]~$\approx$~200 mg L$^{-1}$). 

To disperse the rods in an apolar solvent like cyclohexane the silica shell was made hydrophobic by coating it with octadecyltrimethoxysilane (OTMS). To this end, the silica-coated AuNR dispersion (750 $\mu$L) was diluted with ethanol (1.75 mL) to which OTMS (250 $\mu$L) and butylamine (125 $\mu$L) were added. The mixture was sonicated for 2 h at 30-40 $\degree$C.
Thereafter, the reaction mixture was centrifuged at low speed (100 g for 5 min), washed with toluene, centrifuged at 7,000 g for 10 min, washed twice with cyclohexane (2.0 mL) and redispersed in cyclohexane (250 $\mu$L, [Au]~$\approx$~600 mg L$^{-1}$). 

The spherical SiO$_2$@AuNR supraparticles were made \itshape via \upshape emulsification of an apolar particle dispersion in a larger polar phase~\cite{Besseling2015}. The polar phase consisted of dextran (400 mg) and sodium dodecyl sulfate (SDS) (50 mg) dissolved in H$_2$O (10 mL). The apolar phase consisted of cyclohexane (200 $\mu$L) containing OTMS-functionalised silica-coated AuNRs ([Au]~$\approx$~600 mg L$^{-1}$). The emulsification was done by shortly pre-mixing the apolar and polar phase in a vortex shaker after which it was placed in a sonication bath for 1 minute. Afterwards, the vial was covered with parafilm containing several small holes and the cyclohexane droplets in the emulsion were slowly dried overnight by shaking in an orbital shaker (IKA KS260 basic). The resulting particles assemblies were collected with centrifugation (500 g for 15 min), washed with H$_2$O (8 and 2 mL), and redispersed in H$_2$O (500 $\mu$L).

\subsection*{Synthesis of colloidal silica assemblies}
Monodisperse 531 nm core-shell silica colloids with gold and fluorescent cores were synthesized. 15 nm gold cores were grown using the inverse sodium citrate reduction method~\cite{OjeaJimenez2011,Fokkema2018}: HAuCl$_4$ (3.4 mL, 25 m\textsc{m}) was added  to a boiling solution of sodium citrate in water (345 mL, 1.0 m\textsc{m}) under constant vigorous stirring. After 15 minutes, water (155 mL) and sodium citrate solution (5 mL, 2.2 m\textsc{m}) were added to the obtained deep red solution. After reheating and boiling for an additional 10 minutes the solution was cooled down to 90~$\degree$C. Growth of the seeds to 30 nm was performed in four steps using a kinetically controlled seeded growth procedure~\cite{Bastus2011}: for every growth step sodium citrate (1.7 mL, 120 m\textsc{m}) and HAuCl$_4$ (1.7 mL, 50 m\textsc{m}) were added followed by 60 minutes stirring at 90 $\degree$C. 100 mL of the obtained solution of gold nanoparticles was functionalized with polyvinylpyrrolidone (PVP)~\cite{Graf2003,Fokkema2018} by the addition of PVP (5 mL, 10 m\textsc{m}, $M_w = 10,000$ g mol$^{-1}$) and 16 hours stirring. The functionalized particles were transferred to ethanol by centrifugation (10 min, 15,000 g) followed by redispersion in ethanol (100 mL).
 
Fluorescent rhodamine B labeled cores with a diameter of $\sim$45 nm were synthesized using a reverse micro-emulsion method~\cite{Osseo-Asare1990}. Rhodamine B isothiocyanate (RITC) was coupled to (3-amino\-propyl)\-tri\-ethoxy\-silane (APTES) prior to the synthesis by mixing RITC (6.0 mg), absolute ethanol (500 $\mu$L) and APTES (12.0 $\mu$L) and stirring for 5 hours. The reverse micro emulsion was prepared by mixing cyclohexane (50 mL), Igepal CO-520 (6.5 mL), tetraethyl orthosilicate (TEOS, 400 $\mu$L) and fluorophore-APTES complex (50 $\mu$L). Particle growth was initiated by the addition of ammonia (750 $\mu$L) and after homogenization the solution was stored for 24 hours. The cyclohexane was removed by rotary evaporation under reduced pressure and the obtained pink viscous liquid was diluted in dimethylformamide (10 mL) and ethanol (10 mL) to obtain a clear pink solution.
 
Next, in two separate reactions, the gold and fluorescent cores were coated with a non-fluorescent silica to obtain a total diameter of $\sim$200 nm using a seeded growth procedure based on the St\"{o}ber method~\cite{Giesche1994}. After cleaning \itshape via \upshape repeated centrifugation and redispersion in ethanol, the weight fractions of both solutions were determined, which were used to prepare a 1 to 100 (gold to fluorescent core) mixture in ethanol. Further silica growth was performed to obtain particles with a total diameter 531 nm (<2\% polydispersity index (PDI), 100 particles, transmission electron microscopy), after cleaning by repeated centrifugation and redispersion in ethanol to remove small silica spheres caused by secondary nucleation.

396 nm monodisperse core-shell silica colloids with a fluorescent core were synthesized in the following way. First, using a reverse microemulsion method, a silica core of about $\sim$50 nm was synthesized~\cite{Osseo-Asare1990}. Next, using the seeded St\"{o}ber growth method~\cite{Giesche1994}, a fluorescein isothiocyanate doped silica shell was grown around the core to a diameter of $\sim$200 nm, followed by two silica shells without dye, arriving at a total diameter of 396 nm (1\% PDI, static light scattering).

For the assembly of a colloidal crystal of the 531 nm silica particles, an adaption at elevated temperature of the method by Jiang \textit{et al.}~\cite{Jiang1999} was used to speed up the evaporation process. A cover glass (\#1.5H) was placed under a small angle of $\sim$5$\degree$ in a particle in ethanol dispersion (8 mL, 1 v\%) inside a 20 mL vial. Together with a 100 mL beaker filled with ethanol the vial was placed in a 50 $\degree$C preheated oven (RS-IF-203 Incufridge, Revolutionary Science) and covered with a large beaker placed upside down. After 16 hours the cover glass was removed from the dispersion and a crystal had formed on the cover glass. Any particles sticking to the back of the cover glass were removed by wiping it with an ethanol soaked tissue.

\subsection*{Transmission electron tomography}
The transmission electron microscopy (TEM) tomography was performed on a FEI Talos F200X operated at 200\,kV in STEM-HAADF (scanning transmission electron microscopy - high angle annular dark field) imaging mode. A droplet of aqueous dispersion containing the AuNR assemblies was dried on a special tomography copper grid with parallel bars and a R2/2 Quantifoil film (Electron Microscopy Sciences). The tomography grid was placed in a high tilt holder (Fischione, FP90997/19 tomography holder). The sample was tilted from -70 to +70$\degree$ with a tilt step of 2$\degree$. The tilt images were recorded with 2048 $\times$ 2048 pixels per image, 0.24 nm per pixel, a dwell time per pixel of 1.40 $\mu$s and a total frame time of 6.37 s. The camera length of the HAADF-STEM detector was set to 160 mm. The probe current was 40 pA. Data processing, comprising alignment of the tilt-images \itshape via \upshape cross-correlation and subsequent reconstruction using a simultaneous iterative reconstruction technique (SIRT) algorithm (100 iterations), was carried out in TomoJ (version 2.31)~\cite{TOMOJ}.

\subsection*{FIB-SEM tomography}
The AuNR assemblies dispersion was drop casted on silicon wafer, which was placed on top of an aluminium SEM stub and connected with a conductive carbon tape. The colloidal crystal and binary glass were first infiltrated with a resin to fill the air pockets between the particles. To this end, the colloidal crystal and binary glass were embedded in resin (Lowicryl HM20) and cured overnight in an oven at 65$\degree$C. The cover glasses with the colloidal crystal and the binary glass were attached on an aluminum SEM stub with carbon tape. To prevent charging of the samples under the electron beam, a conductive pathway was created by bridging the top of the cover glass and the stub with a strip of carbon tape. Additionally, the colloidal crystal and binary glass were coated with a 5 nm thick layer of platinum, using a Cressington HQ280 sputter coater. 

The FIB-SEM tomography of the AuNR assembly was performed in a Helios Nanolab G3 UC FIB-SEM (Thermo Fisher Scientific) under high-vacuum conditions (10$^{-6}$ mbar). \textit{In situ} Pt deposition ($\sim$100 nm thick) was accomplished across an AuNR supraparticle by ion beam induced deposition prior to the tomography routine.  Subsequently, the FIB (30 kV, 7.7 pA) milled 160 consecutive slices with a width of 2.5 $\mu$m and a nominal slice thickness of 3 nm. The SEM (2 kV, 100 pA) recorded images in SE and BSE mode (Ultra-High Resolution mode) with a scan resolution of 2304~$\times$~2048 pixels per image, 0.324~$\times$~0.411 nm per pixel and dwell time 3 $\mu$s per pixel. 

FIB-SEM tomography of the colloidal crystal was performed in a Scios FIB-SEM (Thermo Fisher Scientific). Standard preparation procedures (Pt deposition, milling of trenches and polishing of the cross section to be imaged) were performed manually prior to the execution of the tomography routine. The FIB (30 kV, 300 pA) milled 212 consecutive slices with a width of 22 $\mu$m, a calculated depth of 20 $\mu$m and a nominal slice thickness of 50 nm. The SEM (3.5 kV, 100 pA) recorded images (3072 $\times$ 2048 pixels, pixel size 10 nm, dwell time 6 $\mu$s) with the T1 detector in BSE mode. 

FIB-SEM tomography of the binary glass was also performed in a Scios FIB-SEM (Thermo Fisher Scientific). Again, standard preparation procedures were performed manually. Following, the FIB (30 kV,  300 pA) milled 100 consecutive slices with a width of 35 $\mu$m, a calculated depth of 15 $\mu$m and a nominal slice thickness of 50 nm. The SEM (3.5 kV, 100 pA) recorded images (3072 $\times$ 2048 pixels, pixel size 9.4 nm, dwell time 6 $\mu$s) with the T1 detector in BSE mode.

\subsection*{Confocal microscopy}
For confocal microscopy imaging, the binary glass was index matched with a glycerol/$n$-butanol mixture ($n_D^{23}=1.44$). A Leica TCS SP8 confocal microscope equipped with a super continuum white light laser (SuperK, NKT Photonics), a HyD detector and a 100$\times$/1.4 NA confocal objective was used to image the glass. The sample was sequentially scanned with the pinhole set to 1 airy unit to, first, image the rhodamine B dyed particles with the excitation laser set to 550 nm and the detection range from 565 to 687 nm and, second, the FITC dyed particles with the excitation laser set to 488 nm and the detection range from 498 to 590 nm. The voxel size was 31 $\times$ 31 $\times$ 50 nm$^3$ ($X \times Y \times Z$).

\subsection*{Deconvolution}
The confocal image stack was deconvoluted with a theoretical point spread function using the classic maximum likelihood estimation restoration method in the Huygens software (17.04, Scientific Volume Imaging) to a final signal-to-noise ratio of 20.

\subsection*{Particle identification}
To find the positions and orientations of the rods we used the algorithm as described by Besseling \emph{et al}.~\cite{Besseling2015} 
We colored the rods depending on their orientation with $c_\mathrm{red}=|n_x|$, $c_\mathrm{green}=1/2-n_y/2$ and $c_\mathrm{blue}=1/2-n_z/2$ 
where $n_x, n_y$ and $n_z$ are the components of the normalized orientation vector $\mathbf{n}$ along the length of the rod.

To determine the positions of the spherical particles in the FIB-SEM datasets we used a new algorithm of which we will give short description
here. A schematic overview of the main steps in the gradient based tracking method is given in Figure S1. 

After alignment and an initial filtering step the images were blurred with a Gaussian blur (typically 
with $d=1.0$ pixels) to remove noise. Next, the gradients 
of the image were calculated in the $x$, $y$ and $z$ directions resulting in 3 bitmaps ($G_x, G_y, G_z$) 
containing both negative as well as positive values. We also produced a kernel from an ideal image containing a single particle with the 
same dimensions as the particles that we want to locate and blurred this by the same amount. 
We then calculated its gradients in 3D ($K_x, K_y, K_z$) and the convolution (by FFT) of 
the gradient images with the kernel $G_x * K_x + G_y * K_y + G_z * K_z$,
this final image can be seen as Hough transform~\cite{Illingworth1988} and produces
a sharp peak at the location of each particle. 
We then found all local maxima in this image brighter than a predetermined threshold and fitted their position 
with a quadratic function to obtain sub-pixel accuracy. 
For the binary sample we used several (typically 10) kernels for particles with an increasing diameter and 
searched for a maximum in the resulting 4D dataset (\textit{x}, \textit{y}, \textit{z} and diameter).
The distribution of sizes was fitted with two Gaussians and the intersection of the two (475 nm) was chosen to distinguish the small and large particles in the assembly. 

The positions of the particles in the confocal data sets were determined after
image restoration using an extension to 3D~\cite{Blaaderen1995} of a classic 2D tracking algorithm~\cite{Crocker1996}. 

\subsection*{Quantitative analysis}
Radial distribution functions were calculated in the following way from the coordinates of the particles. First, a histogram of the distances between all pairs of $N_{exp}$ particles was calculated. Next, a box, determined by the minimum and maximum values of the coordinates in all three dimensions, was filled with $N_{ig}$ ideal gas particles, of which also a pair distance histogram was calculated. The experimental histogram was divided by the ideal gas histogram, and if $N_{ig} \neq N_{exp}$ the distribution was normalized by a factor of $(\frac{N_{ig}}{N_{exp}})^2$.

From the coordinates of the particles obtained by FIB-SEM tomography the crystal structure was identified using bond orientational order parameters~\cite{TenWolde1996, Lechner2008}. First, a set of numbers was calculated for every particle, based on spherical harmonics $Y_{lm}$:
\begin{equation}
q_{lm} (i) = \frac{1}{n_c(i)} \sum\limits_{j=1}^{n_c(i)} Y_{lm} ( \uvec{r}_{ij}),
\end{equation}
where $n_c(i)$ is the number of nearest neighbors of particle $i$, $l$ an integer (in our case 4 or 6), $m$ an integer running from $-l$ to $l$ and $\uvec{r}_{ij}$ the unit vector pointing from particle $i$ to particle $j$. The nearest neighbours are defined as the particles within cut-off distance $r_c$ from particle $i$. This cut-off was determined from the first minimum of the radial distribution function $g(r)$, corresponding to $r_c \approx 1.4 d$, where $d$ is the particle diameter.
Next, the particles are considered crystalline or liquid using the Ten Wolde criterion~\cite{TenWolde1996}. The correlation between the $q_{lm} (i)$ of every particle with the $q_{lm} (j)$ values of its neighbors was calculated:
\begin{equation}
c_{l}(ij) = \frac{ \sum\limits_{m=-l}^{l} q_{lm} (i) q_{lm}^{*} (j) }{\sqrt{\sum\limits_{m=-l}^{l} |q_{lm} (i)|^2 } \sqrt{\sum\limits_{m=-l}^{l} |q_{lm} (j)|^2} }, 
\end{equation}
where $q_{lm}^{*} (j)$ is the complex conjugate of $q_{lm} (j)$. The neighbors $j$ of each particle $i$ were considered connected when $c_{l}(ij) > 0.6$ and the particle $i$ was considered crystalline when the amount of connected neighbors exceeded $7$. Since hexagonal order was expected we chose $l=6$ to distinguish crystalline and liquid particles.

Next, the crystalline particles were classified having face-centered cubic (FCC) or hexagonal close-packed (HCP) order using the $\bar{w}_l$ order parameter~\cite{Lechner2008}. To calculate this, first the $q_{lm}$ set of numbers of particle $i$ is averaged with the values of its neighbors:
\begin{equation}
\bar{q}_{lm} (i) = \frac{1}{N_c(i)} \sum\limits_{k=0}^{N_c(i)} q_{lm}(k),
\end{equation}
where $N_c(i)$ is the number neighbors $n_c(i)$ of particle $i$ plus itself.  This set of numbers then yields the rotationally invariant averaged local bond orientational order parameter:
\begin{equation}
\bar{w}_l (i) = \frac{\sum\limits_{m_1+m_2+m_3=0}
\begin{pmatrix}
l & l & l \\
m_1 & m_2 & m_3
\end{pmatrix}
\bar{q}_{lm_1}(i) \bar{q}_{lm_2}(i) \bar{q}_{lm_3}(i)}{\left( \sum\limits_{m=-l}^{l}|\bar{q}_{lm} (i)|^2 \right) ^{3/2}},
\end{equation}
where $\begin{psmallmatrix}
l & l & l \\
m_1 & m_2 & m_3
\end{psmallmatrix}$ is the Wigner 3-$j$ symbol and the integers $m_1$, $m_2$ and $m_3$ run from $-l$ to $+l$, but are limited to the case where $m_1+m_2+m_3=0$.
The particles are classified as FCC-like when $\bar{w}_4 < 0$ and HCP-like when $\bar{w}_4 > 0$.


\section*{Acknowledgements}
The authors are grateful to J. Fermie for the help with FIB-SEM sample preparation and H. Meeldijk for EM support. The authors thank T.-S. Deng for useful discussion on the AuNR synthesis and self-assembly. This project has received funding from the European Research Council (ERC) under the European Union's Horizon 2020 research and innovation programme (ERC-2014-CoG No 648991) and the ERC under the European Unions Seventh Framework Programme (FP-2007-2013)/ERC Advanced Grant Agreement \#291667 HierarSACol. JvdH also acknowledges the Graduate programme of the Debye Institute for Nanomaterials Science (Utrecht University), which is facilitated by the grant 022.004.016 of the NWO, the Netherlands Organisation for Scientific research. MH was supported by the Netherlands Center for Multiscale Catalytic Energy Conversion (MCEC).

\section*{Author Contributions}
AvB initiated the project.
JvdH was supervised by PdJ and AvB.
EvdW, DdW, MH and MB were supervised by AvB. 
YL was supervised by MvH.
JF was supervised by HG. 
MB, JF and AvB synthesized the particles.
JF and EvdW prepared the colloidal crystal.
MB prepared the supraparticles under supervision of JvdH.
YL and DdW performed FIB-SEM tomography measurements.
JvdH performed the TEM tomography measurements.
YL and JvdH carried out the the TEM tomography data analysis.
MH performed particle identification of FIB-SEM and TEM tomograms.
EvdW performed the confocal microscopy measurements, particle identification and analysis.
JvdH, EvdW and AvB co-wrote the paper.
All authors analyzed and discussed the results.

\bibliography{MainText_FIB-SEM} 
\bibliographystyle{rsc} 

\end{document}